\documentclass[aps,groupedaddress,twocolumn,floats]{revtex4}
\usepackage{amsmath,amsfonts,graphicx}

\begin{document}

\title{Attractive Potential around a Thermionically Emitting Microparticle}
\author{G.~L. Delzanno$^{1, 2}$, G. Lapenta$^{1, 3}$, M. Rosenberg$^{4}$}
\affiliation{$^1$ Istituto Nazionale per la Fisica della Materia (INFM), Politecnico di Torino, Italy. \\
$^2$ Burning Plasma Research Group, Dipartimento di Energetica, Politecnico di Torino, Italy. \\
$^3$ Plasma Theory Group, Theoretical Division, Los Alamos National Laboratory,
 Los Alamos NM 87545, USA. \\ $^4$ Department of Physics, University of San Diego,  San
Diego, CA 92110, USA. }

\begin{abstract}
We present a simulation study of the charging  of a  dust grain immersed in a plasma, considering the
effect of thermionic electron emission from the grain. It is shown that the OML theory is no longer
reliable when electron emission becomes large: screening can no longer be treated within the
Debye-Huckel approach  and an attractive potential well can form, leading to the possibility of
attractive forces on other grains with the same polarity. We suggest to perform laboratory experiments
where emitting dust grains could be used to create non-conventional dust crystals or macro-molecules.
\end{abstract}

\maketitle

{\bf Introduction:} The study of the charging of objects immersed in a plasma is a classic problem of
plasma physics with many applications~\cite{shukla}, ranging from space problems to dusty plasmas to
probe theory for plasma diagnostics. Recently, the link between condensed matter or high
energy-density plasma and strongly coupled dusty plasmas~\cite{morfil} has renewed the interest in the
process of charging and shielding of dust in plasmas.

The interaction between a plasma and an object is due to the plasma particles that hit the object
surface and are captured. In absence of other processes, the higher mobility of the electrons leads to
more electron captures by the object, which tends to charge negatively. However, in certain
conditions, other processes need to be considered. For example, if the object immersed in the plasma
is sufficiently warm, a significant number of electrons can be emitted by thermionic effect, altering
the balance between electron and ion captures, reducing the negative charge on the object, or even
reversing its sign. An example of this process is given by small objects entering the Earth's
atmosphere (meteoroids). Recent work has shown that the heating of the meteoroids, due to their
interaction with the atmosphere, can produce a considerable thermionic emission which can lead to
positively charged meteoroids~\cite{Sorasio}.

In the present work, we will consider how electron emission changes the process of charging of an
object immersed in a plasma, considering self-consistently charge collection on the object and the
screening by the surrounding plasma. We focus particularly on the process of thermionic emission, but
the results also apply to the similar cases of photoemission and secondary emission that create a
current of electrons emitted  by the object. Two primary conclusions are reached. First, the process
of electron emission by the object reduces the charges as expected by the orbit motion limited (OML)
theory~\cite{shukla}. However, the quantitative effect of the thermionic emission predicted by the OML
theory is accurate only for small objects. We find that, for objects larger than the Debye length, the
OML becomes grossly inaccurate. Second, in the presence of thermionic emission when the object is
charged positively, the screening potential develops an attractive well. In contrast with the typical
monotonic behavior predicted by the Debye-Huckel theory, we observe a potential well due to the
presence of an excess of electron charge trapped around the emitting object.

Looking at the literature, we have found experimental evidence that heated emissive probes can
determine a non-monotonic behaviour of the plasma potential~\cite{intrator}. This is commonly called
the virtual cathode, namely a region of zero electric field associated with a local excess of negative
charges. However, as far as we know, the importance of this mechanism on the charging process
occurring in a dusty plasma and its implications have not yet been recognized. The consequences of
this behaviour of the shielding potential can be considerable since potential wells can provide
regions of attraction for other objects with the same sign of charge. Although the present mechanism
is not the only instance when particles of the same charge immersed in a plasma can attract each other
(see \cite{nambu, shukla1} and references therein), the mechanism presented here can be tested
experimentally. For example photoemission could be used instead of thermionic emission in existing
experiments. Such experiments would be best conducted in microgravity (e.g. on the Alpha space
station) where other attractive mechanisms (e.g. wake field~\cite{nambu, shukla1}, ion flow
alignment~\cite{lapenta03}) may be less important.

{\bf Charging in Presence of Thermionic Emission}: We consider a
spherical, isolated dust grain of radius $a$ immersed in a
neutral, unmagnetized plasma consisting of electrons and singly
charged ions. The grain is stationary, located at $r=0$.
Ion and electron collisions with the neutral gas background are neglected.
Electrons and ions have different masses, $m_e$ and $m_i$, and temperatures,
$T_e$ and $T_i$, respectively. The grain has a surface temperature
$T_d$ and can emit thermionic electrons ($W$ being its work function).
The characteristic lengths of the system are the electron Debye length  $\lambda_{\rm
De}$, the ion Debye length $\lambda_{\rm Di}$ and the linearized
Debye length $\lambda_{\rm Dlin}$, defined as $ 1/\lambda_{\rm
Dlin}^2=1/\lambda_{\rm De}^2+1/\lambda_{\rm Di}^2$.

In the simplest model, neglecting  any emissions from the dust
particle surface, the grain is charged by the surrounding plasma
(primary charging). Initially, electrons are more mobile than ions
and charge the grain negatively by hitting its surface. Hence, the
grain acquires a negative potential and creates an electric field
that repels electrons and attracts ions. A dynamical equilibrium
is eventually reached when the electron current to the dust is
equal to the ion current. The OML theory~\cite{shukla} provides
 a description of the mechanism, and gives the floating potential
 on the dust, $\phi(a)$, as a function of the plasma properties.
 Once $\phi(a)$ is known,
the electric charge on the dust is determined by $Q_{\rm OML}=4 \pi \varepsilon_0 a \left(
1+\displaystyle{{a}/{\lambda_{\rm Dlin}}}\right) \phi(a)$ if one considers a Debye-Huckel potential
around the dust with screening length given by $\lambda_{\rm Dlin}$. Indeed, the OML theory is a good
approximation for thick sheaths where $a \ll \lambda_D$, but breaks down for $a \gg
\lambda_D$~\cite{lapenta2, Daugherty}.

The presence of electron emission from the dust (either
photoelectric or thermionic) affects crucially the potential
distribution around the dust. In this letter we focus on
thermionic emission. The starting point for a theoretical analysis
of the thermionic current is the Sommerfeld model of a metal where
the energy states are uniformly distributed and the free electrons
have a Fermi distribution of probability to occupy a certain
energy state.
%To obtain the number of thermionic electrons
%injected in the system by the dust, we have to distinguish between
We have to distinguish between
positively and negatively charged dust grains. In fact, when the
dust grain is negatively charged, any electron with energy $1/2
m_e v_r^2>\psi$ ($\psi$ being the minumum energy required to
overcome the surface barrier) will be emitted, leading to the
following thermionic current \cite{Sodha}
\begin{equation}
I_{\rm th}=\displaystyle{\frac{16 \pi^2 a^2 e m_e k^2 T_d^2}{h^3}} {\rm exp} \left(-\frac{W}{kT_d} \right),
\label{Ith<0}
\end{equation}
known as the Richardson-Dushman expression. When the grain is positively charged, the situation is
slightly different as the electrons have to overcome the floating potential as well as the surface
barrier. The thermionic current for a positively charged grain is \cite{Sodha}:
%is obtained as \cite{Sodha}:
\begin{equation}
I_{\rm th}=\displaystyle{\frac{16 \pi^2 a^2 e m_e k^2 T_d^2}{h^3}} \left(1+ \frac{e\phi(a)}{kT_d} \right)
{\rm exp} \left(-\frac{W+e\phi(a)}{kT_d} \right).
\label{Ith>0}
\end{equation}
When thermionic emission is added to the OML framework, the
equilibrium floating potential is established by balancing the
ion and electron currents from the surrounding plasma with the
thermionic current emitted by the dust.

{\bf Simulation Method}: To study the charging of a thermionically emitting dust particle, we have
developed a PIC code~\cite{Birdsall} for a spherical plasma with the stationary grain at the center
and an outer radius $R$. The problem under investigation requires special boundary conditions. At the
outer boundary some particles leave the system while others must be injected to represent an infinite
plasma medium outside the simulation domain. The algorithm used to inject the particles is the same
widely used in the literature~\cite{Birdsall}. At the inner boundary, the plasma particles reaching
the grain surface are removed from the simulation and their charge is accumulated to the central dust
grain, affecting its floating potential. The same injection method used for the outer boundary can be
applied also to the thermionic emission at the inner boundary, but using the dust temperature and not
the plasma electron temperature to evaluate the distribution function of the emitted electrons. In PIC
simulations, the emitted electrons are followed accurately and the electrons that cannot overcome the
dust attraction return to the dust; it follows that the emission current must always be computed with
Eq. (\ref{Ith<0}), to avoid counting the retarding potential twice.

We have chosen the parameters of the system according to typical experimental conditions. In
particular, we consider a Maxwellian plasma with electron temperature $T_e=1$ eV,  ion temperature
$T_i=0.2$ eV ($T_e/T_i=5$) and an outer radius of the system $R=500$ $\mu$m. Moreover, the plasma far
away from the dust grain is Maxwellian at rest, with density $n_{\infty}=6 \cdot 10^{15}$ part/m$^3$.
These parameters correspond to the electron Debye length $\lambda_{\rm De}=96.0$ $\mu$m, the ion Debye
length $\lambda_{\rm Di}=42.9$ $\mu$m and the electron plasma frequency $\omega_{\rm pe}=4.37 \cdot
10^{9} s^{-1}$. The linearized Debye length is $\lambda_{\rm Dlin}=39.1$ $\mu$m. The electron mass is
chosen with its physical value, but the ion mass is only 100 times larger. This unphysical choice is
common in the literature and is required to keep the cost of the simulation manageable. All the
simulations are made with an initial number of particles $N_{e}=N_{i}=200000$, located on a uniform
computational grid with $N_{g}=200$ cells. The time step is chosen to satisfy the CFL condition,
$\Delta t=10^{-11}$ s. The radial position and two velocity components (radial and tangential) are
stored for each particle during the simulation.

In the simulations, we start from a uniform Maxwellian plasma and
let the system relax self-consistently until the charge on the
dust grain and the shielding potential around it reach a steady
state. At equilibrium, the dust charge fluctuates due to
collection of plasma particles and we consider the mean value
defined as an average over a time interval of $70\omega_{pe}^{-1}$
s, which is a sufficiently large multiple of the dust charging
time to provide a filter of the high frequency fluctuations.
Hereafter, "time average" will imply an average over the last
$70\omega_{pe}^{-1}$ of the simulation, when the steady state is
fully reached.

Note that all the results presented in the paper are obtained in absence of collisions between plasma
particles and neutrals. Thus our results should be reliable when the mean free path for collisions,
$\lambda_{\rm coll}$, is much greater than the electron Debye length, $\lambda_{\rm De}$. This
requirement is well met for example for weakly ionized plasmas: a glow discharge with pressure $p \sim
10$ mtorr, degree of ionization $\chi \sim 10^{-5}$, density of neutrals $n_{g}\sim 10^{20}$ m$^{-3}$,
plasma density $n \sim 10^{15}$ m$^{-3}$, leads to $\lambda_{\rm coll}/\lambda_{\rm De} \sim 65$.

{\bf Results}: To validate our simulation tool, we consider first
the charging of a dust particle in absence of any emission. We
consider a dust of radius $a=10$ $\mu$m. Since $a/ \lambda_{\rm
Dlin} \simeq 0.2 $, the OML theory should be a good approximation
and we expect our code to agree with theoretical predictions. This
is indeed true, the time average floating charge is $Q_d=-1.54\cdot 10^{-15}$
C, while the one predicted by the OML theory is $Q_{\rm
OML}=-1.64\cdot 10^{-15}$ C. The relative difference is defined as
$|Q_d-Q_{\rm OML}|/\max(Q_d,Q_{\rm OML})=6 \%$. At dynamical equilibrium,
the floating potential is $\phi_d=-1.1386$ V, in good agreement
with the one given by the OML theory, $\phi_{\rm OML}=-1.1776$ V
(relative difference $3 \%$). Furthermore, the shielding potential
follows closely the Debye-Huckel expression with
a screening length equal to the linearized Debye length
$\lambda_{\rm Dlin}$.

We have also considered the primary charging mechanism for a dust
of radius $a=80$ $\mu$m. Since $a/ \lambda_{\rm Dlin}  \simeq 2 $,
we expect the OML theory to be unreliable. In fact, our code gives
$Q_d=-2.04 \cdot 10^{-14}$ C while $Q_{\rm OML}=-3.19 \cdot
10^{-14}$ C with a relative difference of $36 \%$. On the other
hand, the value of the floating charge defined by $Q_{\rm De}=4
\pi \varepsilon_0 a \left( 1+a/\lambda_{\rm De}\right) \phi(a)$ is
a good estimate of $Q_d$ ($Q_{\rm De}=-1.92 \cdot 10^{-14}$ C).
This is a consequence of the fact that when the dust radius grows,
the screening length is determined by the
electrons~\cite{Daugherty}. The profile of the time average
shielding potential follows the Debye-Huckel expression but now
with a screening length equal to the electron Debye length
$\lambda_{\rm De}$, as predicted in Ref.~\cite{Daugherty}. The
time average floating potential obtained by the simulation is
$\phi_d=-1.2044$ V and the relative difference with respect to
$\phi_{\rm OML}=-1.1776$ V is $2 \%$. Thus, the OML theory gives a
good estimate also when $a/ \lambda_{\rm Dlin}  \simeq 2$,
provided that the screening length is determined by the electrons.
The value of $\phi_d$ in the present case is more negative than in
the case of $a=10$ $\mu$m due to the development of an absorption
barrier that diminishes the ion current to the dust. Furthermore,
the sheath is wider, of the order of several linearized Debye
lengths. In summary, our code has confirmed all the
theoretical predictions  from the OML theory regarding non
emitting dust particles.

Next, we include thermionic emissions. We consider a dust at $T_d=0.1$ eV  and with work function
$W=2.2$ eV (representative of some metallic oxides).
%Clearly, high values of $W$ diminish the thermionic current and it
%is easy to check that such parameters lead to a positively charged
%grain according to the OML theory.
These parameters lead to a positively charged grain.
We have performed a number of simulations varying
the dust radius $a$. Here we focus on the case $a=80$ $\mu$m to
point out the most relevant aspects of the role of thermionic
emission in the charging mechanism.  The dynamical equilibrium is
reached in approximately 2 electron plasma periods, being determined
essentially by the electron and thermionic currents.
(The ion dynamics is relevant on a longer time scale, creating
the non-monotonic behaviour of the ion density explained below.)
The equilibrium charge is: $Q_d=8.51 \cdot 10^{-15}$ C, where the OML
theory predicts $Q_{\rm OML}=5.17 \cdot 10^{-15}$ C (based on
expression (\ref{Ith>0}) and $\lambda_{\rm Dlin}$) or $Q_{\rm
De}=3.12\cdot 10^{-15}$ C (based on expression (\ref{Ith>0}) and
$\lambda_{\rm De}$). However, when the thermionic effect is taken
into account a comparison of the floating charge of the simulation
and of the OML theory is no longer correct. In fact, the numerical
factor that defines the floating charge from the floating
potential depends on the potential distribution around the dust
which, when thermionic emission is present, is not well
represented by the Debye-Huckel potential (either with
$\lambda_{\rm Dlin}$ or $\lambda_{\rm De}$). Focusing on the
floating potential, we find $\phi_d=0.1016$ V while $\phi_{\rm
OML}=0.1911$ V. Interestingly, when the thermionic effect is
present, the OML theory does not produce an accurate estimate of
the charging mechanism of large grains (also in cases when, in
absence of thermionic emission, its predictions are acceptable).
Moreover, we have checked that, for small objects, the OML theory
is still reliable when electron emission is included.

\begin{figure}
\centering
\includegraphics[width=70mm,height=50mm]{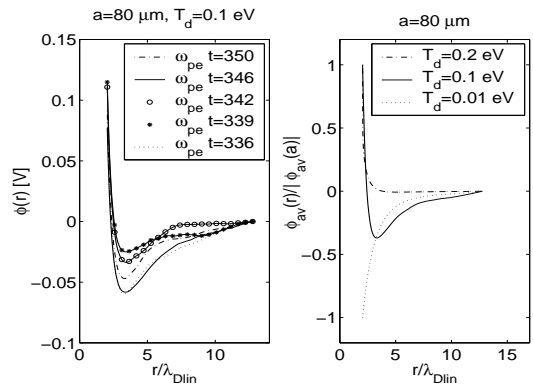}
\caption{Thermionic emission. Shielding potential $\phi(r)$ at
different times (left) and time average shielding potential
(right) as a function of the dust temperature.}
\label{f1}
\end{figure}

Figure~\ref{f1} (left panel) shows the shielding potential at $5$
different times of the simulation.  A potential well is present.
The presence of such well is of considerable interest since it can
lead to attractive forces on another dust particle, even when it
has the same charge. On the right panel of Fig.~\ref{f1} one can
see the time average shielding potential (solid line)
and the potential well is clearly visible. We have also
shown the time average shielding potential obtained by
two simulations with $T_{d}=0.01$ eV (dotted line) and
$T_{d}=0.2$ eV (dash-dotted line).

{\bf Discussion}: How can a potential well form around the dust?
The  explanation comes from the comparison with the case of
absence of thermionic emission.

In presence of sufficient thermionic emission (which is the case
considered above) the dust is positively charged. The resulting
electric field attracts electrons and repels ions. As a
consequence, the electrons emitted from the dust are slowed down
in a region very close to the dust. The more energetic electrons
can escape and contribute to the thermionic current emitted from
the dust but the rest of the thermionic electrons form an electron
cloud. The electron cloud determines an excess of negative charge
and leads to an (equilibrium) potential well.

\begin{figure}
\centering
\includegraphics[width=70mm,height=50mm]{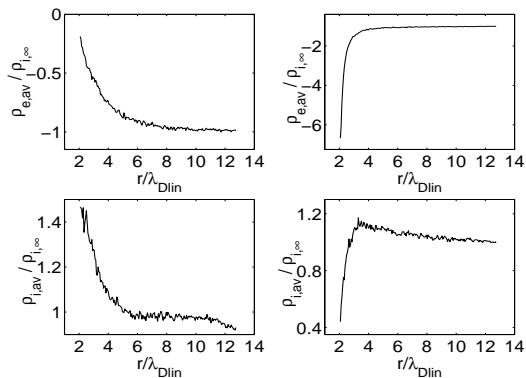}
\caption{Time average normalized electron and ion densities
$\rho_e$ and $\rho_i$ for $a=80$ $\mu$m: primary charging (left)
and thermionic emission (right). } \label{f2}
\end{figure}

To support this explanation,  Fig.~\ref{f2} shows a comparison of
the time average ion and electron density for the case with
only primary charging (left two panels) and with thermionic
emission (right two panels). The densities are normalized with
respect to the unperturbed ion density, $\rho_{i,\infty}$. For the
primary charging case, the densities are perturbed roughly to a
distance of $4$ $\lambda_{\rm Dlin}$ from the dust grain; the
electron density decreases while the ion density increases towards
the grain. The last result is due to spherical geometry and to the
ion angular momentum \cite{Daugherty}. In fact, there are many
ions with high angular momentum that do not strike the grain,
thus leading to an increment of ion density in the sheath
with respect to the equilibrium value.
Consider next the case with thermionic effect (right panels). The
electron density increases close to the grain both due to
thermionic emission and to the attractive potential on the dust.
On the other hand, since the grain is positively charged, the
ion density diminishes approaching it. It can be noticed
that the ion density increases from the dust grain somewhat up to
$4$ $\lambda_{\rm Dlin}$, reaches a maximum and decreases to the
value at rest. Clearly, Fig.~\ref{f2} shows the excess of
electrons needed for the formation of the attractive well observed
above.

\begin{figure}
\centering
\includegraphics[width=70mm,height=50mm]{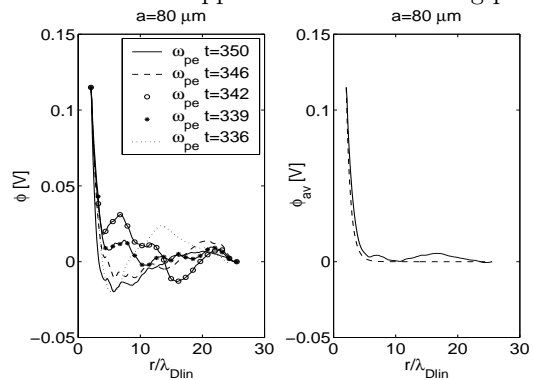}
\caption{Langmuir probe. Shielding potential $\phi(r)$ at different
times (left) and time average shielding potential (right), for a
dust particle with $a=80$ $\mu$m.
The Debye-Huckel potential with $\lambda_{\rm Dlin}$ (dashed line)
is shown for comparison. The potential on the dust is held fixed
at $\phi(a)=0.115$ V and no thermionic electrons are emitted. }
\label{f3}
\end{figure}

We have performed another simulation where we have kept the
potential on the dust fixed at the same potential observed in the
simulation described above, in presence of thermionic emission. In
the present case, we impose the dust potential and we are not
allowing any thermionic emission: only the primary charging
is in effect. Note that this simulation is actually a description of the well known Langmuir probe
used in experimental plasma diagnostics. For this case, Fig. \ref{f3} shows the potential distribution
at different times (left panel) and the time average shielding potential (right panel). One can see
that the  time average potential around the dust is a decreasing monotonic function of radius and
vanishes asymptotically. Clearly, this confirms that the excess of electrons seen in Fig.~\ref{f2}
depends on the thermionic electrons.

Finally, we have found the dust temperature critical for the well
formation, as shown in Fig. \ref{f1} (right). When $T_{d}=0.01$ eV,
the thermionic current is sufficiently small so that the grain is
negatively charged and the shielding potential is an increasing
monotonic function of radius (as for the primary charging case).
When the dust temperature grows, for example up to $T_{d}=0.1$ eV,
an electron cloud forms around the dust and
causes a potential well. However, if $T_{d}$ is increased
further (for example $T_{d}=0.2$ eV), more thermionic electrons
have enough kinetic energy to escape the dust attraction.
At this high dust temperature, the local excess of electrons is
reduced, the potential well disappears and the shielding potential
becomes a monotonically decreasing function of radius.

{\bf Acknowledgments}: Work partially supported by the IGPP-LANL
grant number 03-1217 and by DOE grant DEFG03-97ER54444. G.L.D.
thanks Gianfranco Sorasio for many stimulating
discussions and for his suggestions.

\end{document}